\documentclass{PoS}
\usepackage{amsmath}
\usepackage{lipsum}
\usepackage{enumitem}
\usepackage{soul}

\newcommand{\TeV}{{\ensuremath\rm TeV}}
\newcommand{\GeV}{{\ensuremath\rm GeV}}

\newcommand{\MeV}{{\ensuremath\rm MeV}}
\newcommand{\HS}{\texttt{HiggsSignals}}

\newcommand{\eqn}{equation}
\newcommand{\lam}{\lambda}
\title{IDM and not only}

\ShortTitle{IDM and not only}

\author{Agnieszka Ilnicka\\
        Institute of Physics, University of Z\"urich \\
Winterthurstrasse 190, CH-8057 Z\"urich\\and\\
Physics Department, ETH Z\"urich \\
Otto-Stern-Weg 5, CH - 8093 Z\"urich\\
        E-mail: \email{ailnicka@physik.uzh.ch}}

\author{
\speaker{Maria Krawczyk}\thanks{Supported by part by the National Science Centre, Poland, the HARMONIA project under contract UMO-
2015/18/M/ST2/00518 (2016-2019).}\\
         Faculty of Physics, University of Warsaw\\ ul. Pasteura 5, 02-093 Warsaw, Poland\\
        E-mail: \email{krawczyk@fuw.edu.pl}}

\author{Tania Robens\\
Department of Physics and Astronomy, Michigan State University\\
567 Wilson Road,
East Lansing, MI 48824,
USA\\
E-mail: \email{robens@pa.msu.edu}}

\author{Dorota Soko\l owska \\
 Fcaulty of Physics, University of Warsaw\\  ul. Pasteura 5, 02-093 Warsaw, Poland\\
        E-mail: \email{Dorota.Sokolowska@fuw.edu.pl}
}

\abstract{In this note, we discuss recent updates to a previous scan for the Inert Doublet Model, a two Higgs doublet model with a dark matter candidate. We explicitly show the inclusion of updated constraints on direct detection reduces the allowed parameter space significantly. {We furthermore discuss the IDMS -  an extension of the IDM with a complex singlet.} }

\FullConference{Corfu Summer Institute 2016 "School and Workshops on Elementary Particle Physics and Gravity"\\
		31 August - 23 September, 2016\\
		Corfu, Greece}

\begin{document}
\section{Introduction}
The second run of the LHC is currently in full swing, with more and more data becoming available for an increased center-of-mass energy\footnote{See e.g. \cite{moriond}.}. While the experiments acquire further data to rediscover and establish the scalar sector of the Standard Model {(SM)}, beyond the Standard Model (BSM) scenarios need to be further scrutinized.\\
In this work, we discuss a two Higgs doublet model with a discrete symmetry, which we label $ D$-symmetry,   {with} transformations defined as:

\begin{equation}
\phi_S\to \phi_S, \,\, \phi_D \to - \phi_D, \,\,
\text{SM} \to \text{SM}.
\end{equation}
{This discrete symmetry} 
 is respected by the Lagrangian 
and  the vacuum. The $\phi_S$ doublet plays the same role as the corresponding  doublet in the SM, {providing} the SM-like Higgs particle. This doublet is even under the $D$ symmetry, while the second doublet, $\phi_D$, is $D$-odd. This so called inert or dark doublet contains 4 scalars, two charged and two neutral ones, with the lightest neutral scalar being a natural DM candidate.
This Inert Doublet Model (IDM) was first discussed in \cite{Deshpande:1977rw,Cao:2007rm,Barbieri:2006dq}.  {The model} was studied in context of discovery  {at the} LHC, both for the Higgs boson discovery \cite{Barbieri:2006dq,Cao:2007rm} as well as dark matter discovery e.g. in two lepton + MET channel \cite{Dolle:2009ft,Gustafsson:2012aj}.  Other collider studies and scans of the models parameter space have been presented in \cite{Blinov:2015qva,Ferreira:2015pfi,Diaz:2015pyv,Hashemi:2015swh,Krawczyk:2015xhl,Poulose:2016lvz,Kanemura:2016sos,Datta:2016nfz,deFlorian:2016spz,Hashemi:2016wup,Belyaev:2016lok}. Moreover the model offers rich cosmological phenomenology. {It was {shown} that} the Universe evolution to the current Inert phase may lead through one, two or three phase transitions \cite{Ginzburg:2010wa}, and especially may lead to a {strong enough}  first-order transition, fulfilling one of the Sakharov conditions for the baryogenesis \cite{Hambye:2007vf,Chowdhury:2011ga,Gil:2012ya}. {T}he metastability of the potential was {equally} discussed, {see} \cite{Goudelis:2013uca,Swiezewska:2015paa,Khan:2015ipa}. The dark matter arising from IDM is also referred to as "archetypical WIMP" \cite{LopezHonorez:2006gr}, and thus many astrophysical observations {can be} reinterpreted in its context. 

In this work, we present updates of the scan presented in \cite{Ilnicka:2015jba} (see also \cite{Ilnicka:2015sra,Ilnicka:2015ova}) and especially discuss how they influence the benchmark planes proposed therein. We include updated limits on direct dark matter detection, as well as the total width and branching ratios of the 125 \GeV~ Higgs. Our results should be considered as an intermediate step to a full update of our previous work.

We also present some results on the extension of the IDM by a neutral complex singlet with a complex {vacuum expectation value}. Here,
as in the IDM, the only $D$-odd field in the model is $\phi_D$, while all other fields are $D$-even:
\begin{equation}
 \phi_S \to \phi_S, \;  \phi_D \to - \phi_D, \; \textrm{SM fields} \to  \textrm{SM fields}, \; \chi \to \chi. \label{zzz}
\end{equation}
This IDMS model, in a constrained version  called cIDMS, offers a possibility of a spontaneous violation of CP, {while} at the same time {retaining the} main properties of the IDM, {therefore} being in agreement with LHC and astrophysical data \cite{Bonilla:2014xba}.
\section{The IDM}

The scalar sector of IDM consists of two {{SU(2)}} doublets of complex scalar fields,   $\phi_S$  and $\phi_D$, {{with the $ D$-symmetric potential:}}
\begin{equation}\begin{array}{c}
V=-\frac{1}{2}\left[m_{11}^2(\phi_S^\dagger\phi_S)\!+\! m_{22}^2(\phi_D^\dagger\phi_D)\right]+
\frac{\lambda_1}{2}(\phi_S^\dagger\phi_S)^2\! 
+\!\frac{\lambda_2}{2}(\phi_D^\dagger\phi_D)^2\\[2mm]+\!\lambda_3(\phi_S^\dagger\phi_S)(\phi_D^\dagger\phi_D)\!
\!+\!\lambda_4(\phi_S^\dagger\phi_D)(\phi_D^\dagger\phi_S) +\frac{\lambda_5}{2}\left[(\phi_S^\dagger\phi_D)^2\!
+\!(\phi_D^\dagger\phi_S)^2\right],
\end{array}\label{pot}\end{equation}

Due to this symmetry the  scalar field{s}  {in $\phi_D$} do not mix with the SM-like field {from} $\phi_S$ , and {the} lightest particle of  {the} dark sector is stable.  
{{After {electroweak symmetry breaking} (EWSB), only $\phi_S$ acquires  {a} nonzero vacuum expectation value ($v$).}}
{The dark sector  {contains} 4 new particles: $H$, $A$ and $H^{\pm}$.  {We here choose H to denote the} dark matter (DM) candidate\footnote{ {A priori, any of the new scalars can function as a dark matter candidate. However, we neglect the choice of a charged dark matter candidate, as these are strongly constrained \cite{Chuzhoy:2008zy}. Choosing A instead of H changes the meaning of $\lam_5$, but not the overall phenomenology of the model, cf. \cite{Ilnicka:2015jba}.}}.  \\

 {After EWSB, the model contains a priori seven free parameters. Agreement with the Higgs boson discovery and electroweak precision observables fixes} the SM-like Higgs mass  {$M_h$} and $v$,  {and we are left with} 5 free parameters, which we take {as}
\begin{\eqn}\label{eq:physbas}
M_H, M_A, M_{H^{\pm}}, \lam_2, \lam_{345},
\end{\eqn}
 where  {the} $\lambda$'s correspond coupling{s}  {within the dark sector} and {to the} SM-like Higgs respectively. {In the following, we will use the shortcut}  $\lam_{345} = \lam_{3}+\lam_{4}+\lam_{5}$.

\subsection{Updates on limits on the total width of $h$, branching ratios, and direct dark matter detection}
The work presented here follows previous work   \cite{Ilnicka:2015jba} (see also \cite{Ilnicka:2015sra,Ilnicka:2015ova}) and introduces updates on several of the experimental constraints. In particular
\begin{itemize}
\item{}We use an updated constraint on the total width of the SM-like Higgs $h$, $\Gamma_h\,\leq\,13\,\MeV$ \cite{Khachatryan:2016ctc};
\item{}{We update the}  limits on the invisible decay of $M_h$: we take the results presented in \cite{Khachatryan:2016vau}, leading to $\text{BR}_{h\,\rightarrow\,\text{inv}}\,\leq\,0.16$.
\item{}{We also use} new limits on the branching ratio $h\,\rightarrow\,\gamma\,\gamma$, taken from \cite{Khachatryan:2016vau}, and rendering $\mu\,=\,1.14^{+0.19}_{-0.18}$, which, in combination with the Standard Model value \cite{deFlorian:2016spz} of $\text{BR}\,(h\,\rightarrow\,\gamma\,\gamma)\,=\,2.270\,\times\,10^{-3}$, {{gives}}
\begin{\eqn}\label{eq:rgaganew}
\text{BR}\,(h\,\rightarrow\,\gamma\,\gamma)\,\in\,\left[1.77;3.45\right]\times\,10^{-3}
\end{\eqn}
on a two-sigma level, {{leading to slightly more stringent results than }}the previous constraint \cite{Ilnicka:2015jba}
\begin{\eqn*}
\text{BR}^{\text{old}}\,(h\,\rightarrow\,\gamma\,\gamma)\,\in\,\left[1.76;3.84\right]\times\,10^{-3}.
\end{\eqn*}

\item{}We also implement an update on direct detection bounds, presented in \cite{Akerib:2016vxi}.
\item{} {We confront the model agains the newest experimental bounds as implemented in \texttt{HiggsBounds-5.0.0} {\cite{Bechtle:2008jh, Bechtle:2011sb, Bechtle:2013wla}} and \texttt{HiggsSignals-2.0.0} {\cite{Bechtle:2013xfa}}. \footnote{{Available from http://higgsbounds.hepforge.org/downloads.html.} Note that current release{s are} beta version{s}, and thus should be treated with care.}}

\end{itemize}
 {We plan to extend our scan in the future, e.g. in the direction presented in \cite{Belyaev:2016lok}. }  In this note, we focus on how the above bounds limit the parameter space and benchmark planes presented in \cite{Ilnicka:2015jba} (see also \cite{deFlorian:2016spz}). All other theoretical and experimental constraints follow our scan in \cite{Ilnicka:2015jba}, and we refer to this work for {more details}.

\subsubsection{Results}
The improved constraint on the total width of $M_h$ cited above does not have any direct consequence for the allowed parameter space. While it is important in an intermediate step, we found this always to be superseded by constraints from Higgs signal rates, which we implemented using \HS. The same holds for the updated value on the branching ratio to invisible particles. The update on $h\,\rightarrow\,\gamma\,\gamma$ decay rates leads to slightly stronger constraints for $M_H\,\leq\,M_h$\footnote{Note that in the previous study, we included constraints using \HS, which corresponds to a global fit over all signal strength measurements. Limits obtained with a cut on one of these only might therefore render slightly different results due to the difference in statistical treatment.}. {Afterwards the surviving points were also confronted with the most updates results, including 13 \TeV~analysis as implemented in \texttt{HiggsBounds-5.0.0} and \texttt{HiggsSignals-2.0.0}, but showed no additional impact on the {allowed} parameter space.}\\
On the other hand, the inclusion of new LUX bounds \cite{Akerib:2016vxi} significantly reduces the allowed parameter space. For illustration, we show the benchmark planes before {{(\cite{Ilnicka:2015jba})}} and after the new constraint has been applied. We do this exemplarily for $HA$ and $H\,H^\pm$ production cross sections at a 13 \TeV\,  LHC in Figs.~\ref{fig:compHA} and \ref{fig:compHPH}, respectively. {Implementation of the new direct detection bounds leads to a significant reduction {of valid points} in the $(M_H;\lam_{345})$ plane, while still rendering similar orders of magnitude for the respective production cross sections.}
\begin{figure}[h]
\begin{center}
\includegraphics[width=0.48\textwidth]{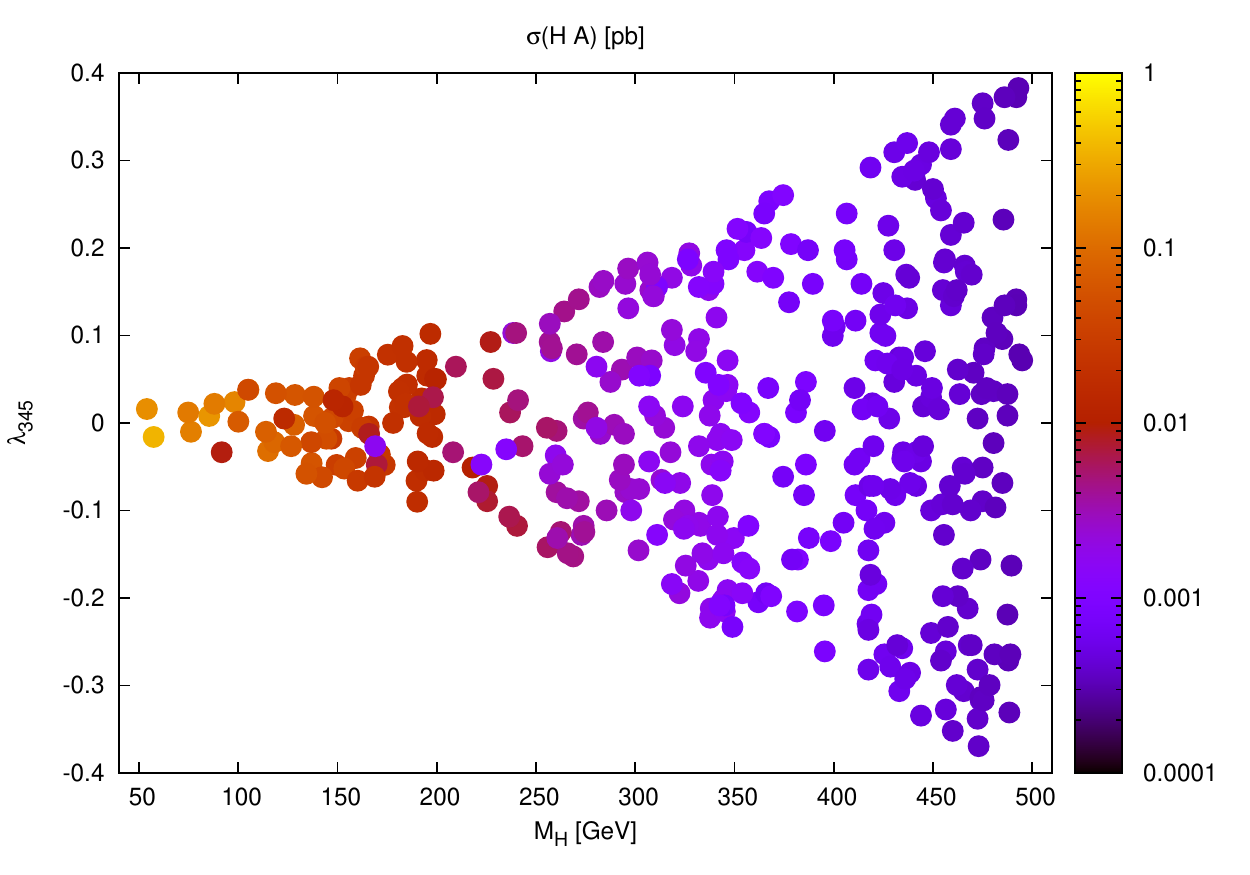}
\includegraphics[width=0.48\textwidth]{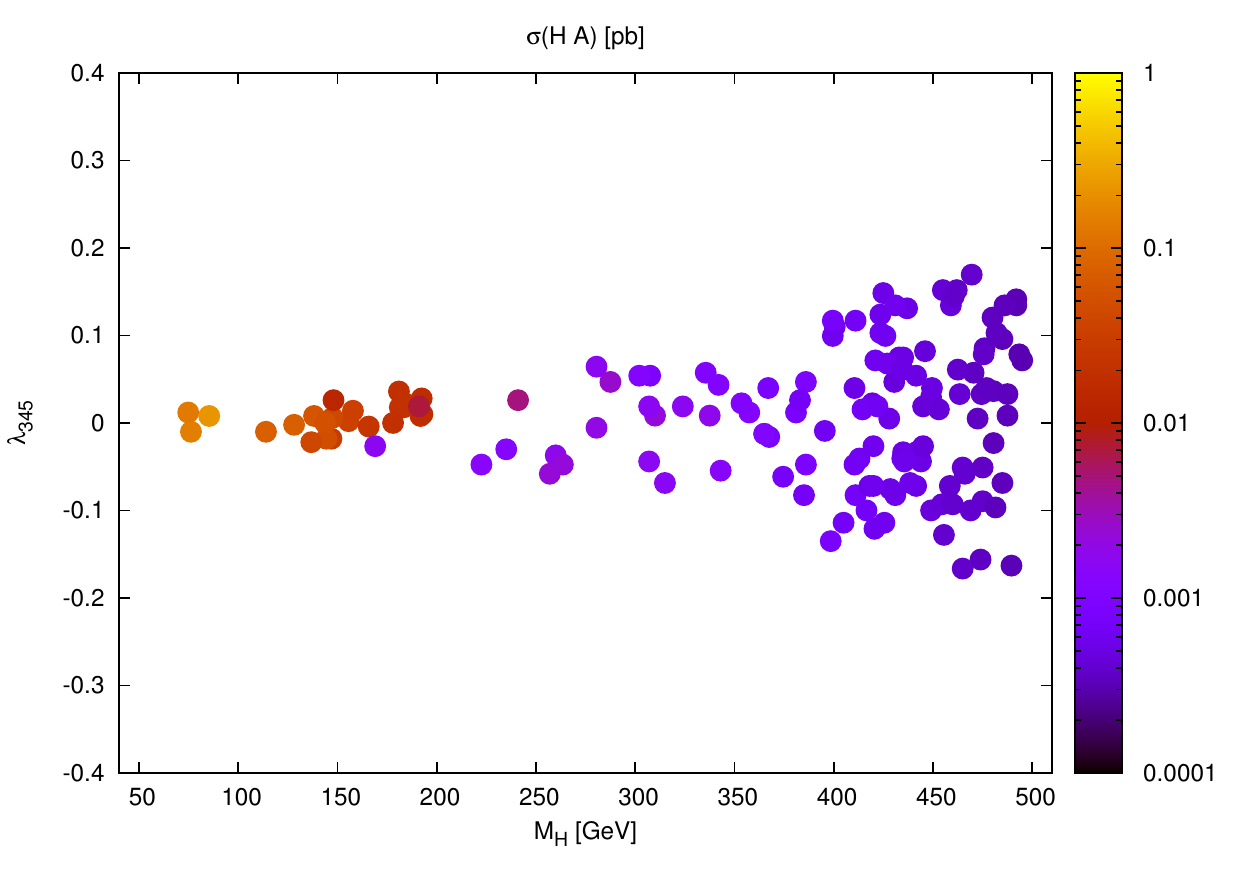}

\includegraphics[width=0.48\textwidth]{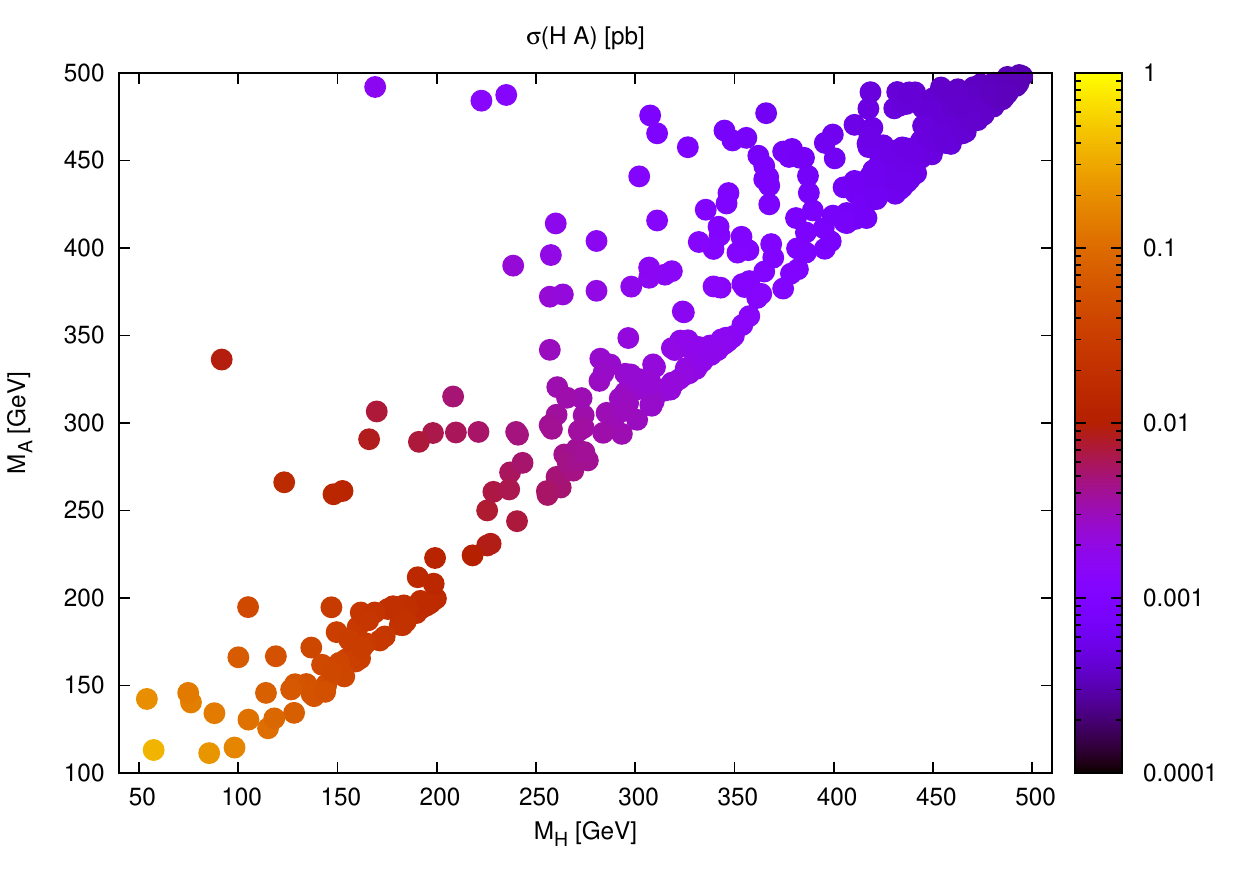}
\includegraphics[width=0.48\textwidth]{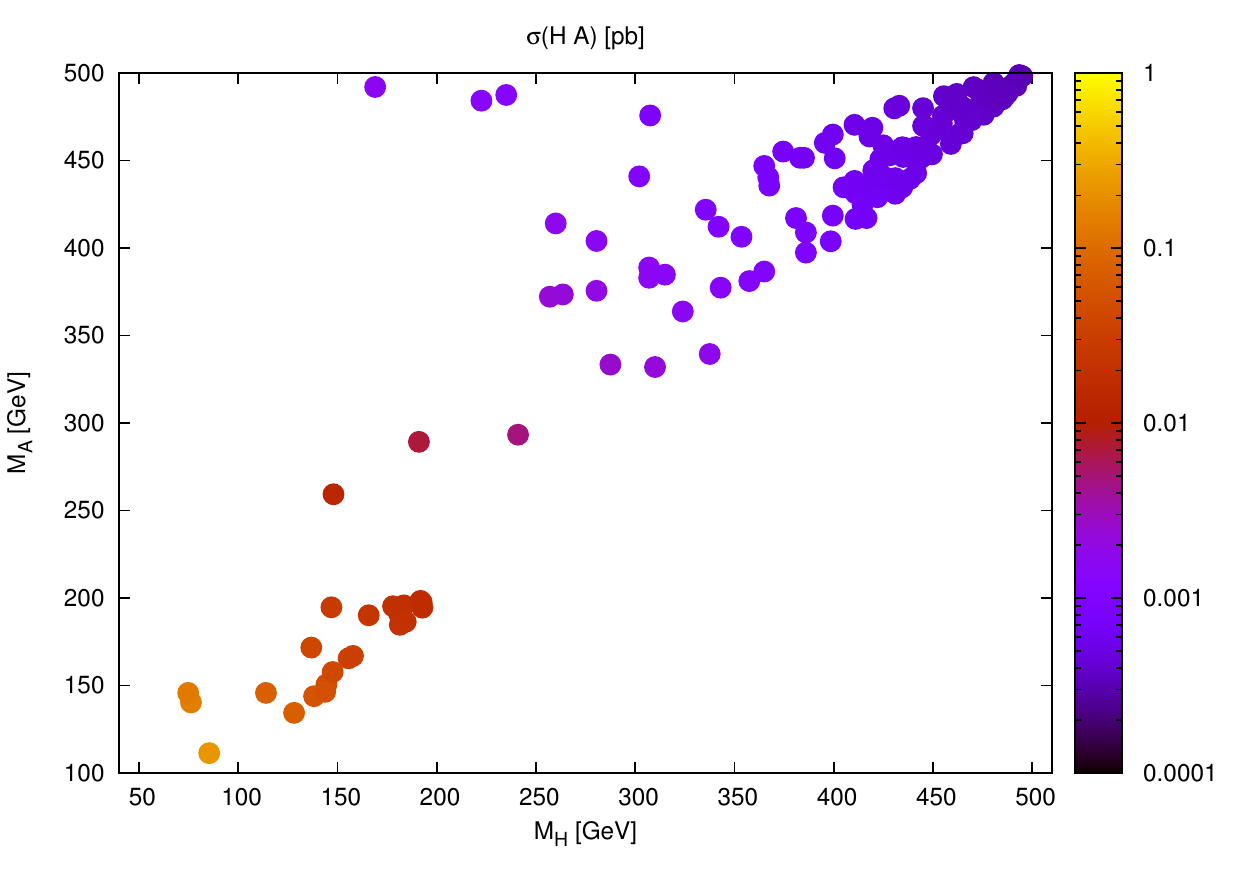}
\caption{\label{fig:compHA} Total production cross section at a 13 \TeV~collider for $HA$ final states, using old \cite{Akerib:2013tjd} {\sl (left)} and new \cite{Akerib:2016vxi} {\sl (right)} limits from direct detection, in the $(M_H,\,\lam_{345})$ {\sl (upper row)} and $(M_H,\,M_A)$ {\sl (lower row)} plane. The new bounds constrain the parameter space significantly.}
\end{center}
\end{figure}
\begin{figure}[h]
\begin{center}
\includegraphics[width=0.48\textwidth]{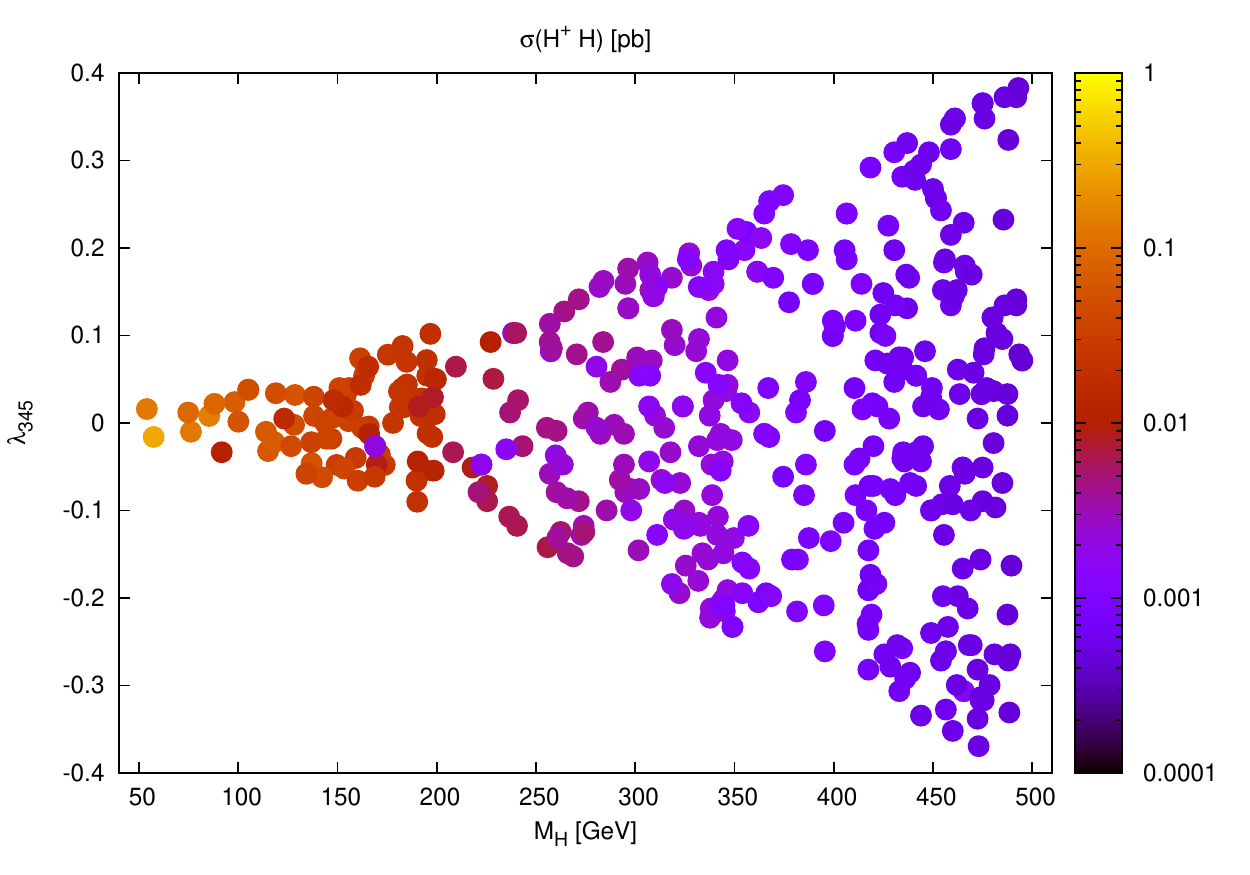}
\includegraphics[width=0.48\textwidth]{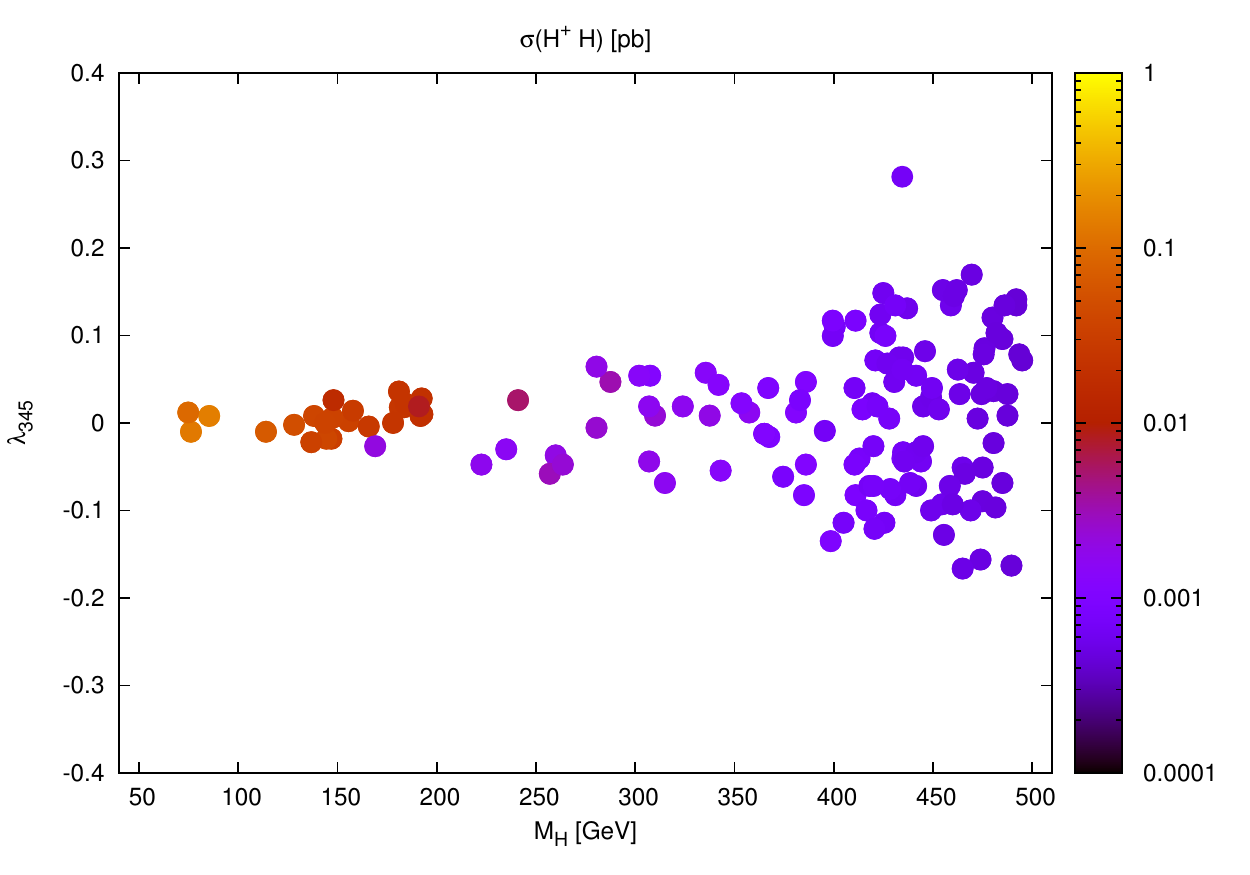}

\includegraphics[width=0.48\textwidth]{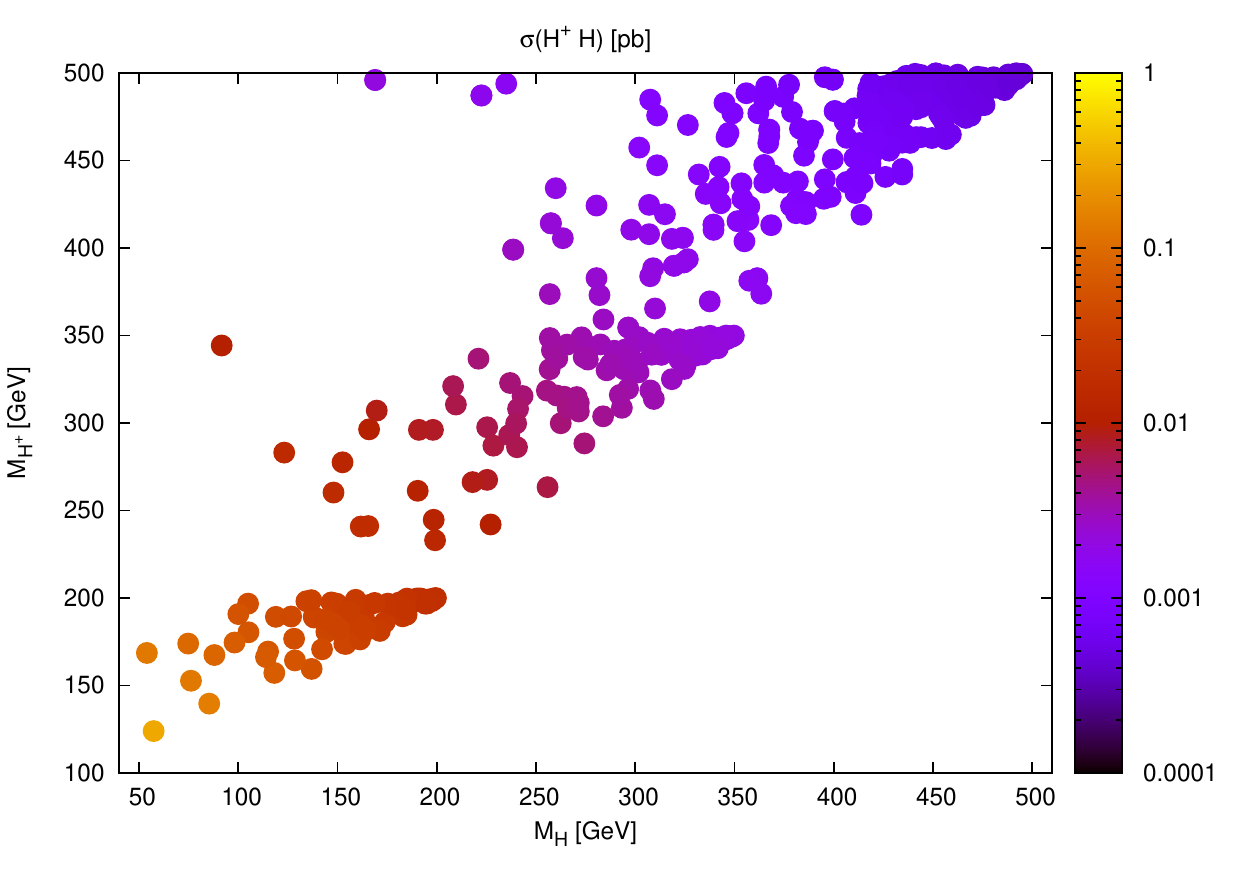}
\includegraphics[width=0.48\textwidth]{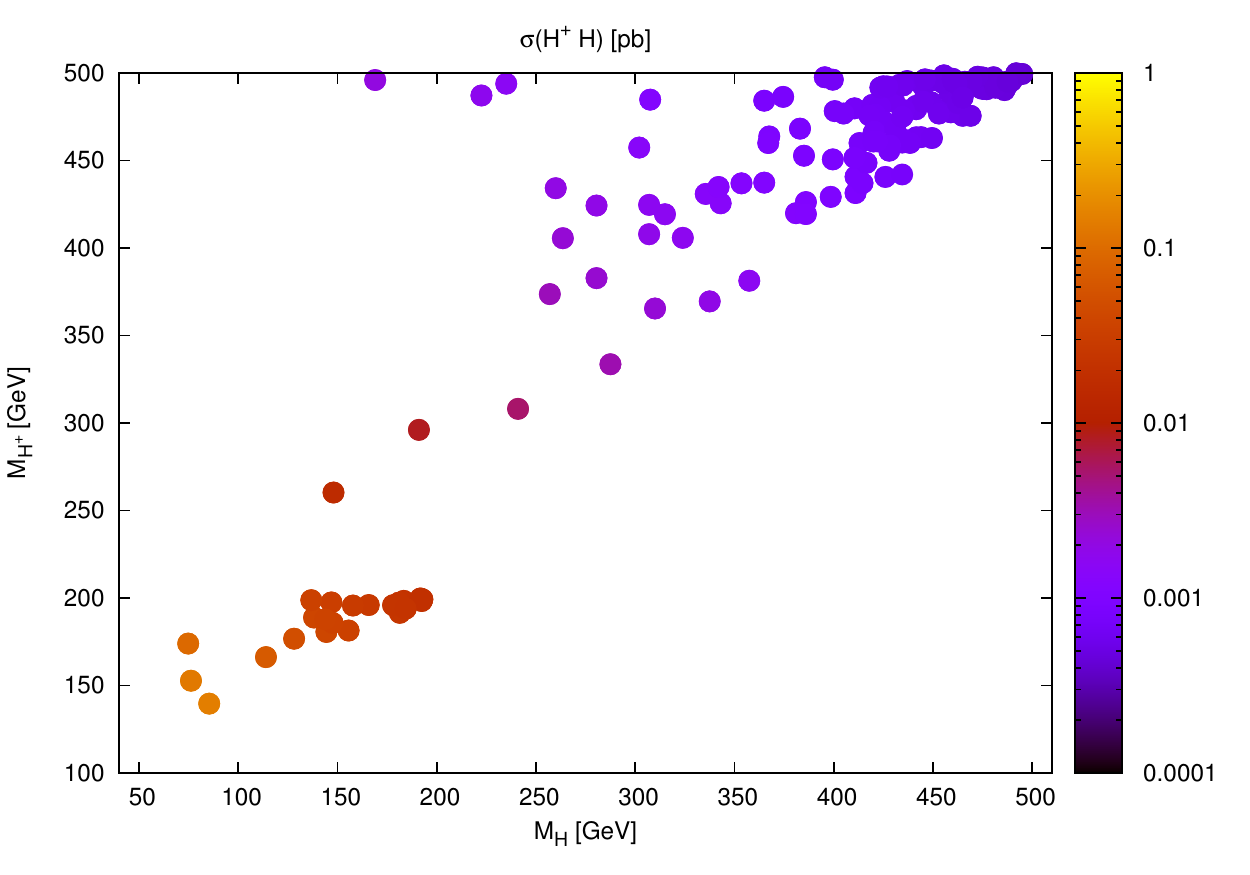}
\caption{\label{fig:compHPH} Total production cross section at a 13 \TeV~collider for $H\,H^\pm$ final states, using old \cite{Akerib:2013tjd} {\sl (left)} and new \cite{Akerib:2016vxi} {\sl (right)} limits from direct detection, in the $(M_H,\,\lam_{345})$ {\sl (upper row)} and $(M_H,\,M_{H^\pm})$ {\sl (lower row)} plane. The new bounds constrain the parameter space significantly.}
\end{center}
\end{figure}
For the case where $M_H\,\leq\,M_h$, the updates on bounds from direct detection mainly play a role in the region where $M_H\,\geq\,M_h/2$, see Fig.~\ref{fig:lowcomp}. {In this case, however, the new limits do not diminish the available parameter space {significantly}.}

\begin{figure}[h]
\begin{center}
\includegraphics[width=0.48\textwidth]{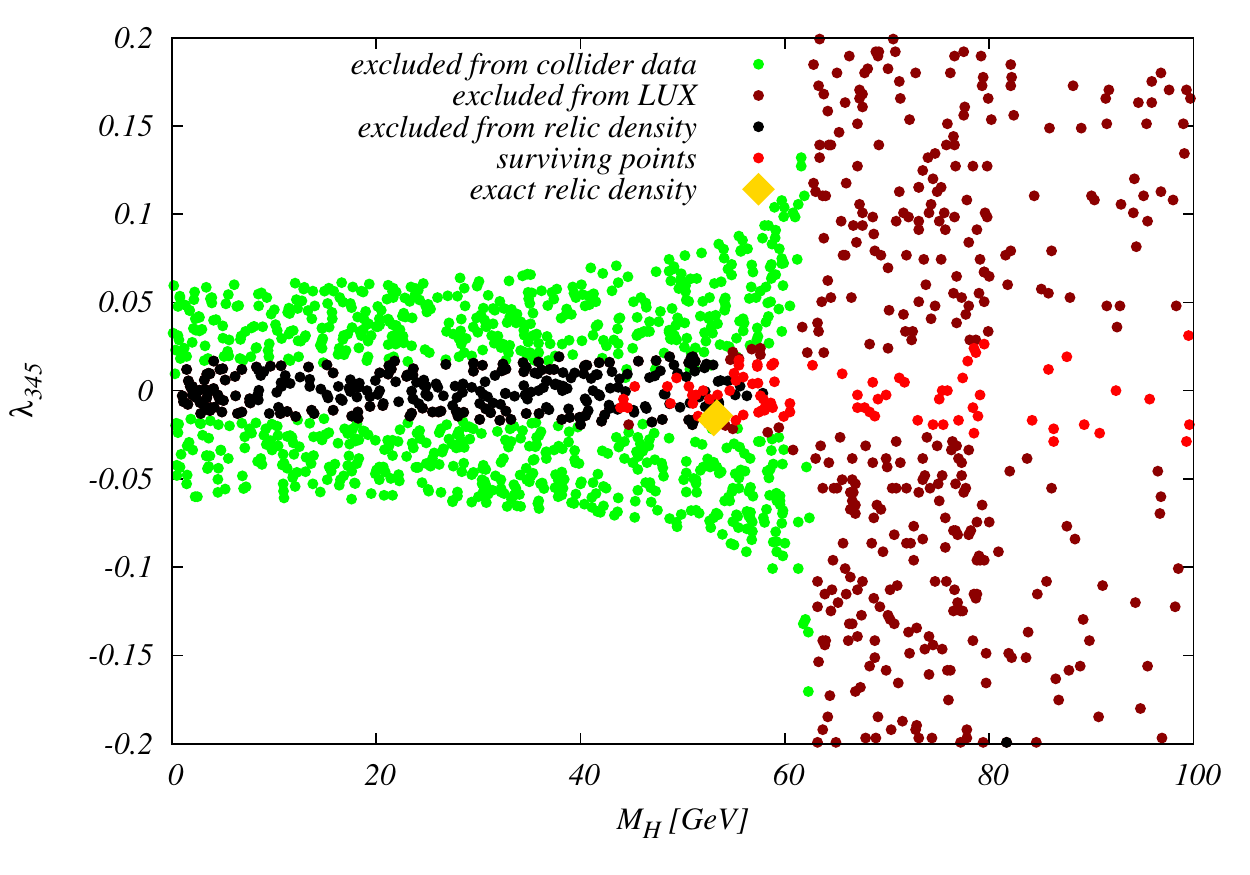}
\includegraphics[width=0.48\textwidth]{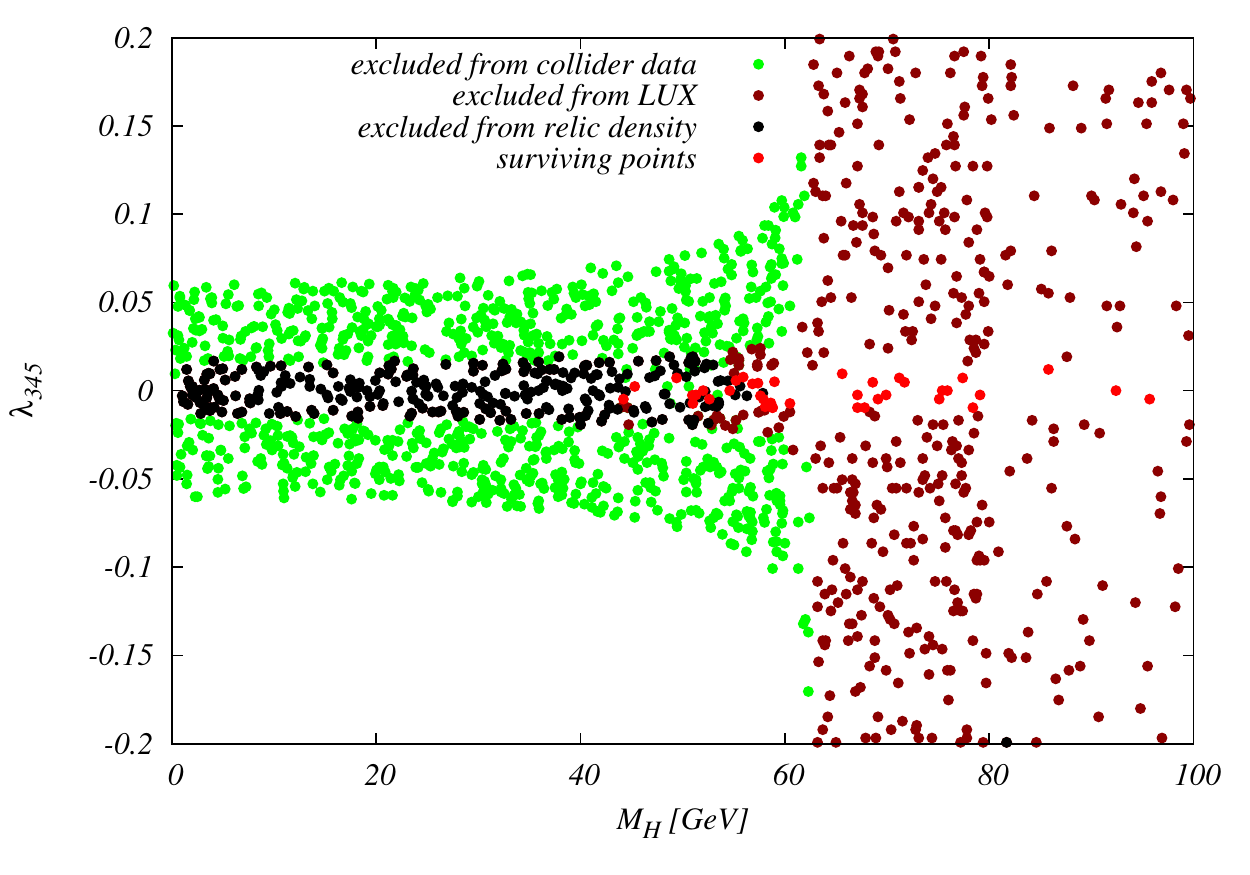}

\caption{\label{fig:lowcomp} Regions in the $(M_H,\lam_{345})$ plane that are allowed {\sl (red)} after all constraints, as well as point excluded from dark matter relic density {\sl (black)}, direct detection {\sl (brown)}, as well as collider data {\sl (green)}. We again display old \cite{Akerib:2013tjd} {\sl (left)} and new \cite{Akerib:2016vxi} {\sl (right)} limits from direct detection. We find minimally stronger constraints for $M_H\,\geq\,M_h/2$. The left hand plot includes two points (displayed on top of each other) which render exact relic density (golden). After the new constraints have been applied, these vanish due to $\text{BR}(h\,\rightarrow\,\gamma\,\gamma)$ as well as the new LUX bound, respectively.}
\end{center}
\end{figure}
In \cite{Ilnicka:2015jba}, we also discussed the case of multi-component dark matter scenarios, pointing out that especially the allowed parameter space for $AA$ pair-production is largely enhanced in such a scenario. When we apply a strict lower limit on $\text{BR}(h\,\rightarrow\,\gamma\gamma)$ instead of a softer limit using \HS, we see that the parameter space is significantly reduced in the low $M_A$/ low $\lam_{345}$ plane, leading to the reduction of available production cross sections by about one to two orders of magnitude, cf. Fig.~\ref{fig:aarcomp}. As discussed in our previous work, this limit can directly be related to regions in the $(\lam_3;M_{H^\pm}/M_h)$ plane, the important parameters in the corresponding BSM contribution in the IDM, {see} Fig.~\ref{fig:rats}.
\begin{figure}[h]
\begin{center}
\includegraphics[width=0.48\textwidth]{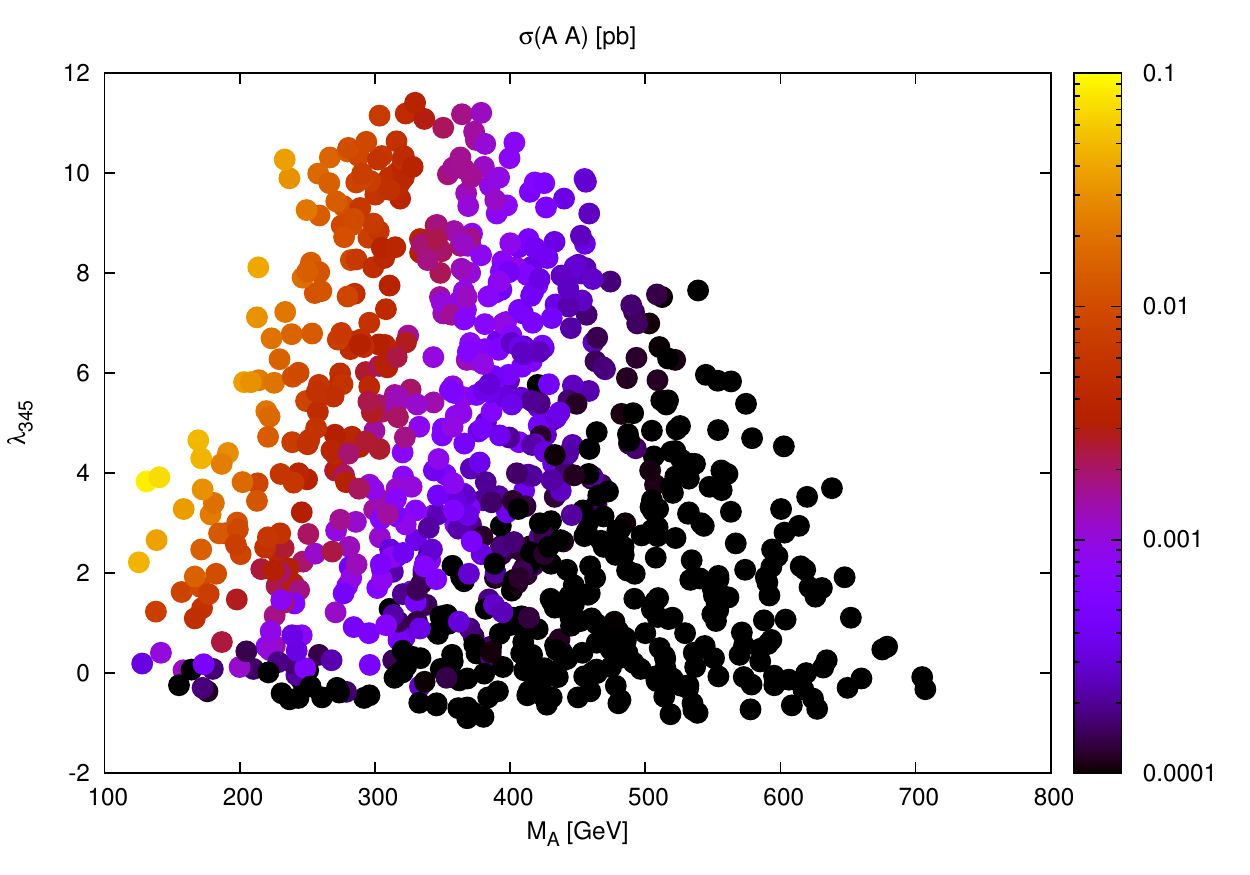}
\includegraphics[width=0.48\textwidth]{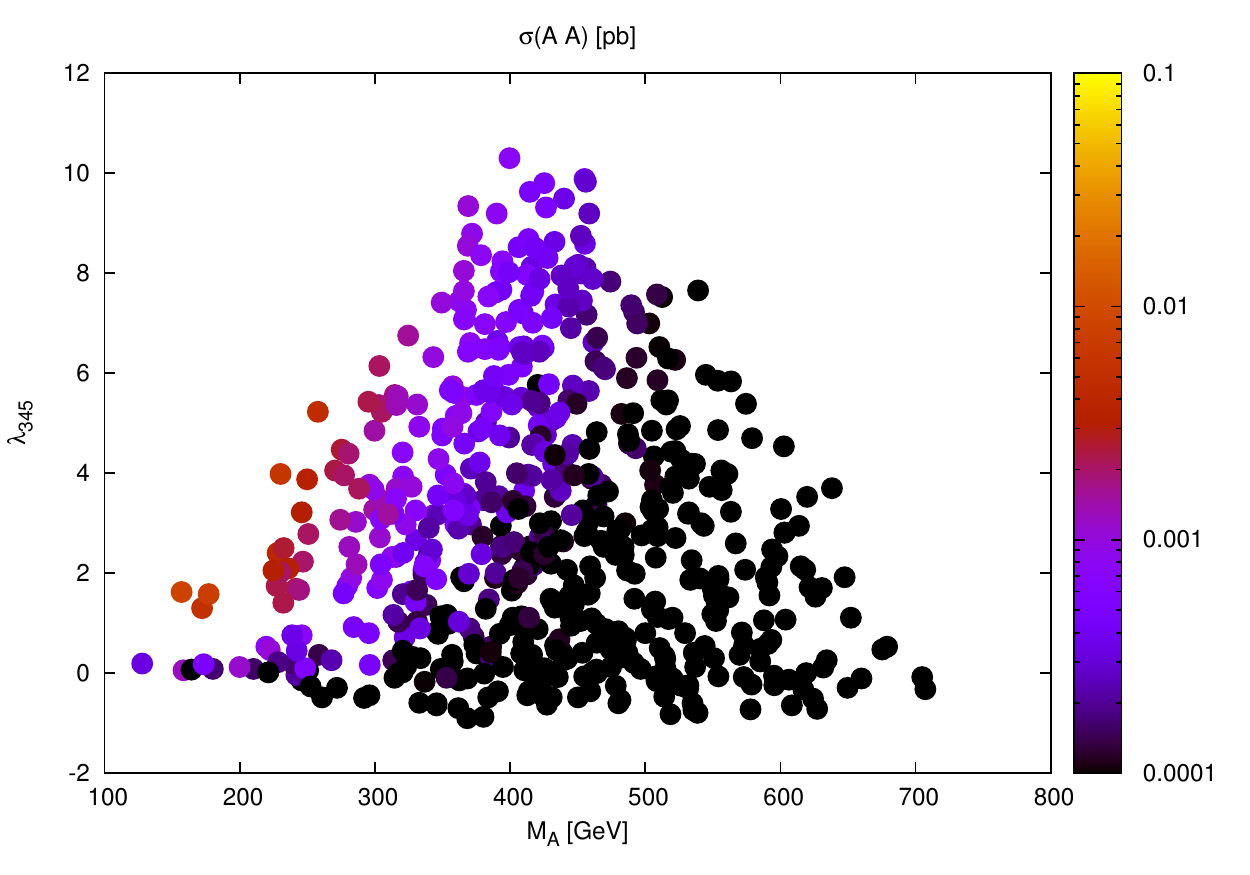}
\caption{\label{fig:aarcomp} Production cross sections for $AA$ final states at a 13 \TeV~ LHC for a multi-component dark matter scenario, as discussed in \cite{Ilnicka:2015jba}. {The left figure is taken from \cite{Ilnicka:2015jba}.} Including a hard cut on $\text{BR}(h\,\rightarrow\,\gamma\,\gamma)$ {{\sl (right)}} reduces the parameter space significantly, leading to a decrease in the maximally allowed production cross sections.}
\end{center}
\end{figure}
\begin{figure}[h]
\begin{center}
\includegraphics[width=0.48\textwidth]{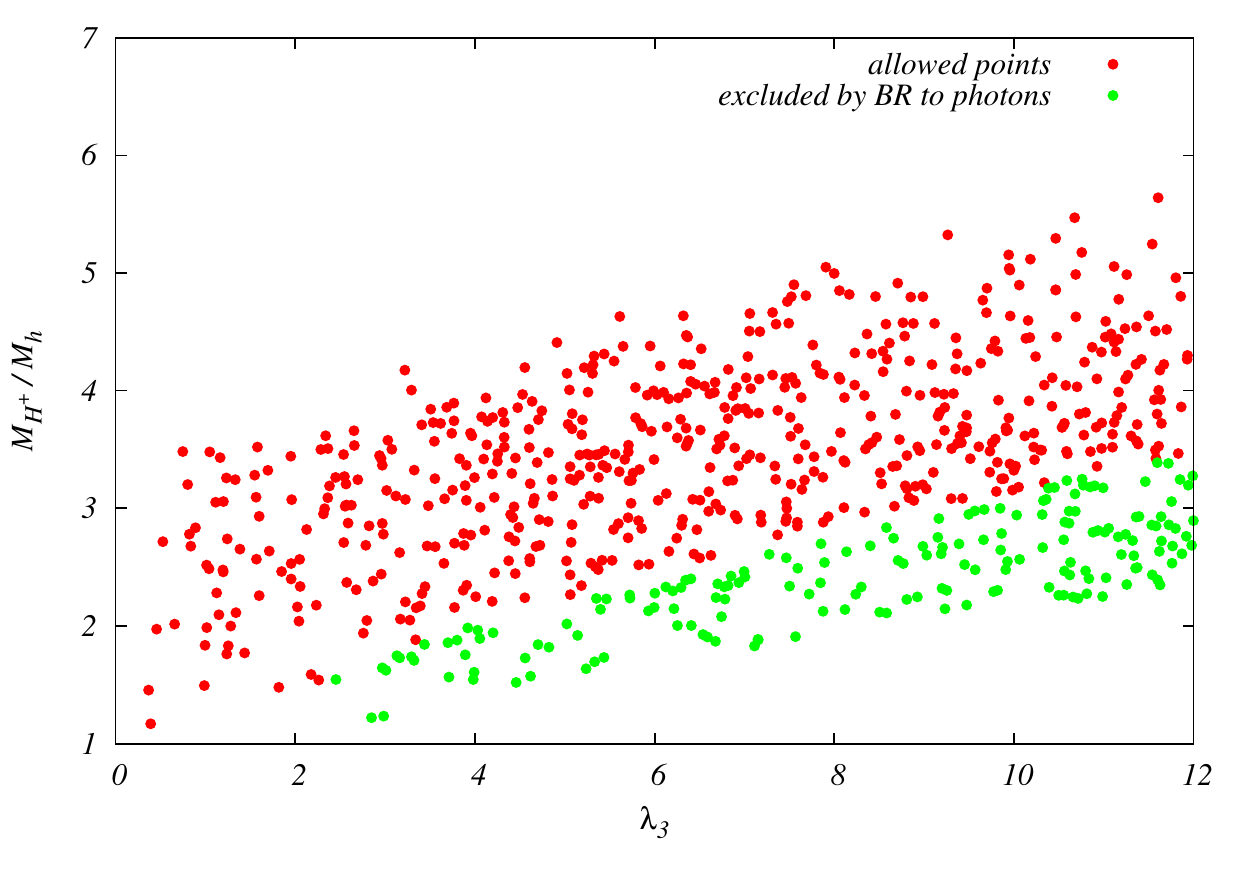}
\includegraphics[width=0.48\textwidth]{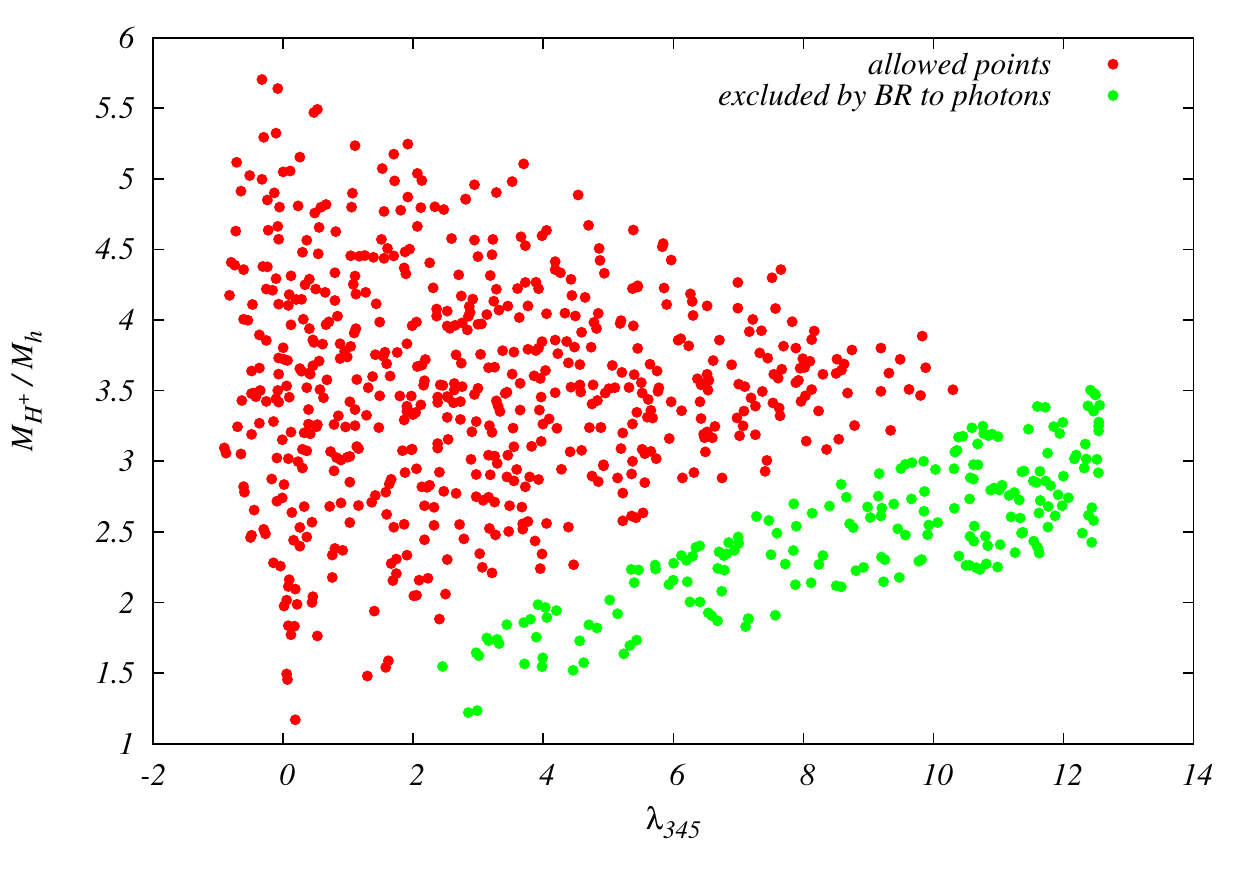}
\caption{\label{fig:rats} Points allowed (red) and excluded (green) the branching ratio $h\,\rightarrow\,\gamma\gamma$, in the $(\lam_3,M_{H^\pm}/M_h)$ {\sl (left)} and $(\lam_{345},M_{H^\pm}/M_h)$ {\sl (right)} plane. We here allow for multicomponent dark matter which significantly opens up the parameter space to large $\lam_{345}$ values.}
\end{center}
\end{figure}


\section{The cIDMS}
The IDM provides a simple, yet compelling candidate for a DM particle, and in general is in agreement with experimental data. However, one significant feature is missing from the IDM: an additional source of CP violation. One possibility of introducing it is by adding a complex field $\chi$ with a complex VEV. This was done in the model called the IDMS: a $D$-symmetric Two-Higgs Doublet Model, the IDM, with a complex singlet \cite{Bonilla:2014xba}. 
As in the IDM, the only $D$-odd field in the model is $\phi_D$, while all other fields are $D$-even:
\begin{equation}
D\;:\; \phi_S \to \phi_S, \;  \phi_D \to - \phi_D, \; \textrm{SM fields} \to  \textrm{SM fields}, \; \chi \to \chi, \label{z2}
\end{equation}
The {constrained version of the model, called  cIDMS, is based on the following  potential}:
\begin{equation} 
    \begin{array}{c}
V = -\frac{1}{2}\left[{m_{11}^2} \phi_S^\dagger\phi_S + {m_{22}^2} \phi_D^\dagger\phi_D \right] 
+ \frac{1}{2}\left[\lambda_1 \left(\phi_S^\dagger\phi_S\right)^2 
+ \lambda_2 \left(\phi_D^\dagger\phi_D \right)^2\right]\\[6mm]
+  \lambda_3 \left(\phi_S^\dagger\phi_S \right) \left(\phi_D^\dagger\phi_D\right) + \lambda_4 
\left(\phi_S^\dagger\phi_D\right) \left(\phi_D^\dagger\phi_S\right) +\frac{\lambda_5}{2}\left[\left(\phi_S^\dagger\phi_D\right)^2\!
+\!\left(\phi_D^\dagger\phi_S\right)^2\right] \\[3mm]
-\frac{m_3^2}{2} \chi^* \chi + \lambda_{s1} (\chi^*\chi)^2 + \Lambda_1(\phi_S^\dagger\phi_S)(\chi^* \chi)\\[2mm]
 -\frac{m_4^2}{2} (\chi^{*2} + \chi^2) + \kappa_2 (\chi^3 + \chi^{*3}) + \kappa_3 [ \chi(\chi^*\chi) + \chi^*(\chi^*\chi)].
    \end{array}
\label{potIDM1S}
\end{equation}
with all parameters real. The first part of the potential corresponds to the doublet interactions from Eq.~(\ref{pot}). New parameters include the mass and interaction terms for the singlet, as well as the portal interaction with the SM-like doublet $\phi_S$ through $\Lambda_1$. 

In this model a $D$-odd scalar doublet provides a good DM candidate (the lightest particle among the $D$-odd $H,A,H^\pm$ -- as in the IDM we chose $H$ to be the lightest), while a complex singlet with a non-zero complex VEV, $\langle \chi \rangle = w e^{i\xi}$, can introduce additional sources of CP violation. The physical Higgs spectrum consists of three scalar particles, $h_1,h_2,h_3$, and one of them ({typically the lightest one, {labelled} $h_1$)} can play {the} role of the SM-like Higgs particle. In that sense the cIDMS can be treated as an extension of the IDM, with many features repeated from the IDM. However, additional Higgs particles can influence Higgs and DM phenomenology.

There are two phenomenolgically interesting cases: (i) only one Higgs particle, playing a role of the SM-like Higgs, is light, while the remaining two are significantly heavier, (ii) where at least one of additional Higgs particles has a mass of the order of $2\,M_{H}$.

\subsection{  {The one and two light Higgs particles scenarios }}

The first scenario, with two additional Higgs particles heavier than the SM-like Higgs and DM particles, resembles the IDM. The low DM mass region, defined as $M_{H} < M_{h_1}/2$, reproduces the known behaviour of Higgs-portal DM models, like the IDM. For $M_{H} \lesssim 53$ GeV it is impossible to fulfil LHC constraints for the Higgs invisible decay branching ratio and relic density measurements at the same time. For $53\,\GeV \lesssim M_H \lesssim 63\,\GeV$  we are in the resonance region of enhanced annihilation {where a} very small {value of the} coupling $\lambda_{345}$ {reproduces the} proper relic density. This region is partially in agreement with 
DM direct detection constraints, however both indirect detection and the LHC results {constrain} the parameter space, as in the IDM. For $M_{H} > M_{h_1}/2$ DM annihilation {into gauge boson final states becomes important}, and the Higgs-DM couplings are generally larger, leading to strong constraints from direct detection experiments. However, as in the IDM, it is still possible to find solutions that are in agreement with all the data.


The most striking change with respect to the IDM arises in the relic density analysis with the possibility 
of having an additional resonance region if the mass of one of additional Higgs particles is smaller than $2\,M_W$. In this case the coupling of SM-like Higgs to DM candidate is very small, therefore it will escape direct detection limits. 

{{ Below (Figs. \ref{fig:idm-idms} and \ref{fig:idmsdet}) we compare these two scenarios for the corresponding two benchmarks denoted  $A1$ and $A4$.
  \begin{itemize}
\item \textrm{\textbf{A1}: }\;$ M_{h_1} = 124.83\,\GeV, \; M_{h_2} = 194.46\,\GeV, \;
M_{h_3} = 239.99\,\GeV, $ 
\item \textrm{\textbf{A4}: }\; $M_{h_1} = 125.36\,\GeV, \; M_{h_2} = 149.89\,\GeV, \;
M_{h_3} = 473.95\,\GeV$.
\end{itemize}}}

\begin{figure}[ht]
\begin{center}
\includegraphics[width=0.47\textwidth]{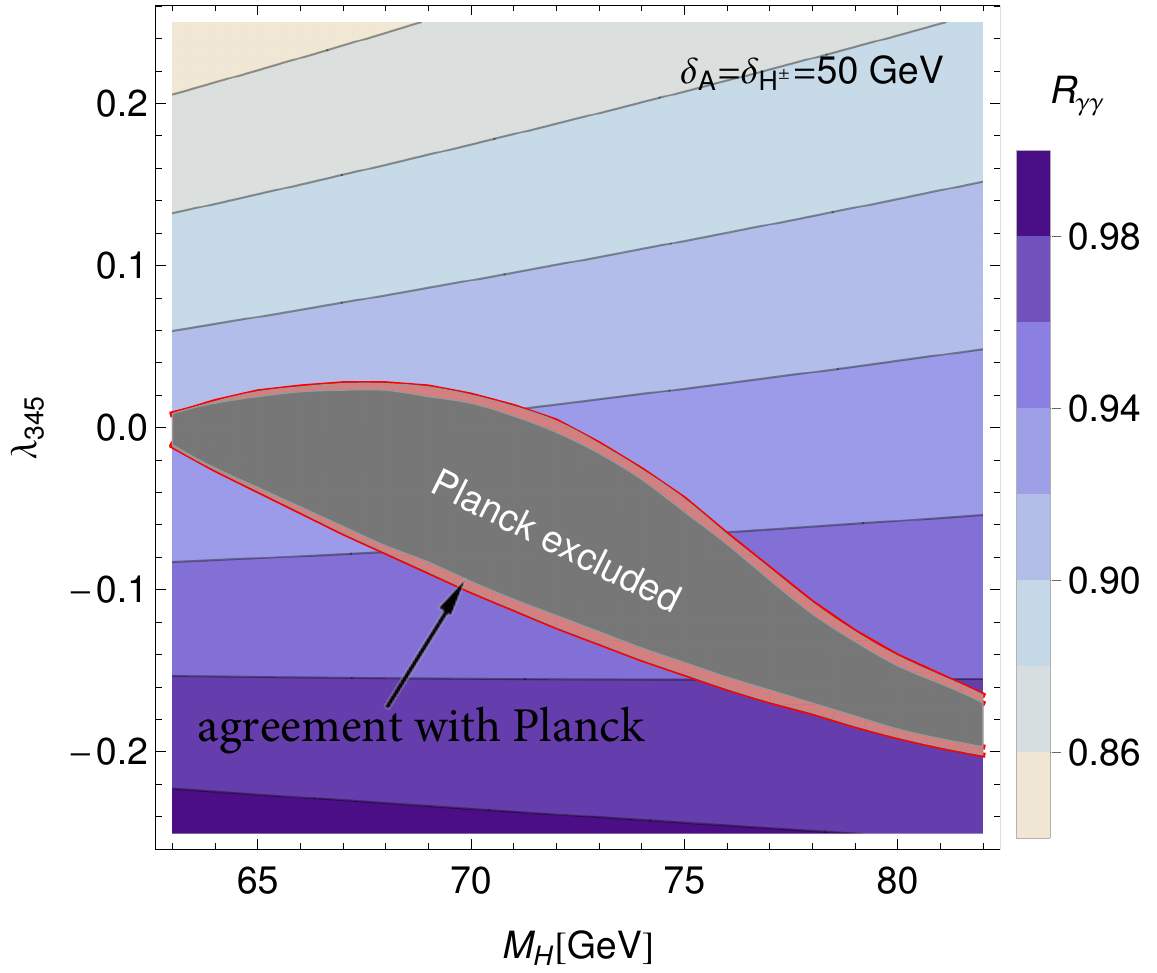}
\includegraphics[width=0.4\textwidth]{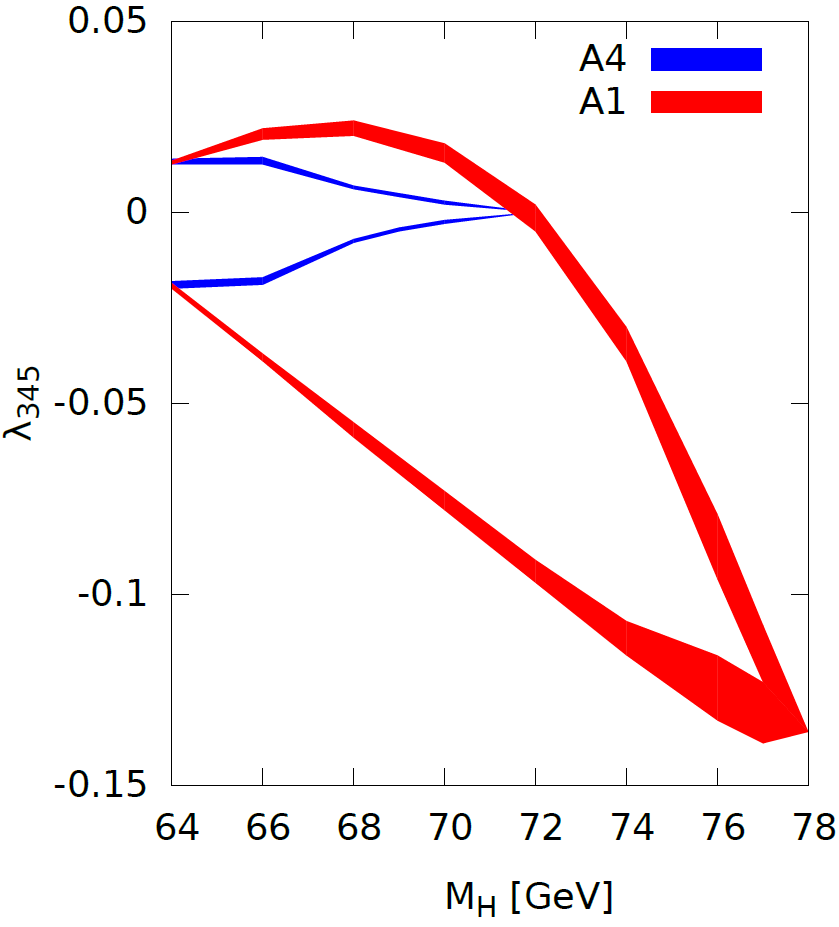}
\caption{\label{fig:idm-idms}  The coupling  $\lambda_{345}$ versus mass of DM particle (H).
{{\sl (Left):}} for the IDM, {{\sl (Right)}} for
cIDMS ($A1$, {\sl red}): case with one light Higgs particle, ($A4$, {\sl blue}): case with
two light Higgs particles. {The red and blue bands constitute regions in agreement with relic density as measured by Planck experiment \cite{Planck:2015xua}}. From \cite{Krawczyk:2015vka} and \cite{Krawczyk:2015xhl}.}
\end{center}
\end{figure}

\begin{figure}[h!]
\begin{center}
\includegraphics[width=0.7\textwidth]{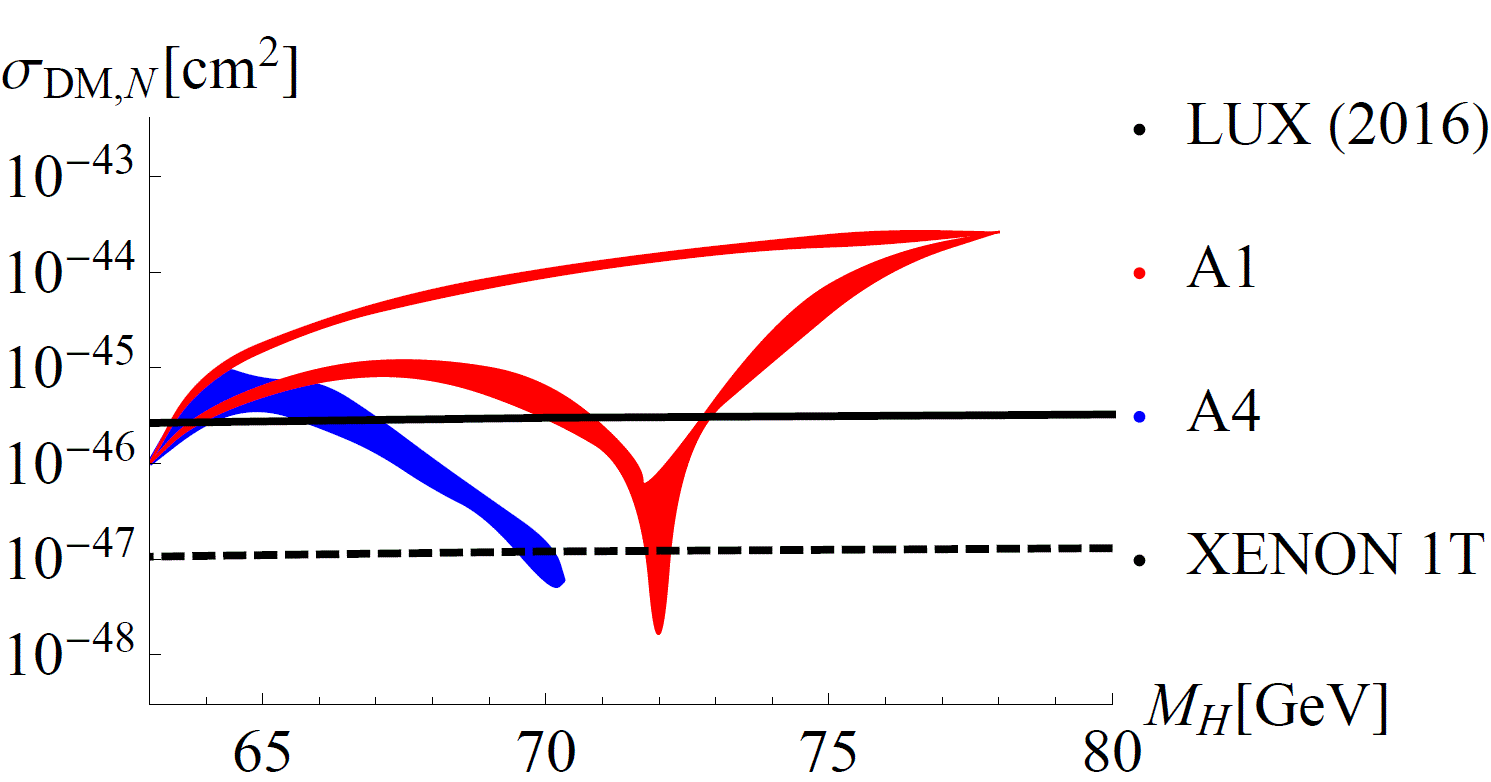}
\caption{\label{fig:idmsdet} DM-nucleon scattering cross-section for the
cIDMS ($A1$, {\sl red}): case with one light Higgs particle, ($A4$, {\sl blue}): case with
two light Higgs particles. {The red and blue bands constitute regions in agreement with relic density as measured by Planck experiment \cite{Planck:2015xua}}. {In addition,} limits from LUX \cite{Akerib:2016vxi} (solid
black) and XENON1T \cite{Aprile:2012zx} (dashed black) are shown. }
\end{center}
\end{figure}

\subsection{LHC constraints}

It is important to stress that in a large part of the parameter space it is the LHC that provides the strongest constraints for the model. If the Higgs-DM coupling is small, we generally escape the direct or indirect detection limits. Therefore, from the point of view of dedicated DM experiments, the small Higgs-DM coupling in the resonance region(s) is favoured. However, even in this region we can expect {significant modifications which influence LHC measurements in the Higgs sector.} 

Strong constraints for the model come particularly from measurements of the {125 \GeV~Higgs decay to $\gamma\gamma$}. In general, {in the cIDMS} this value  is reduced with respect to the SM {prediction $\mu^{\text{SM}}_{h\,\rightarrow\,\gamma\gamma}$}. In the IDM-like case the preferred values {for $h_1$} are {$0.95\,\,{\times}\,\mu^{\text{SM}}_{h\,\rightarrow\,\gamma\gamma}$}. In cases where there are two relatively light Higgs particles we obtain much smaller values of $h_1\to \gamma\gamma$ and $h_1 \to Z\gamma$ signal strengths. These {can differ up to 20 \% from}  the SM value, and -- while not being yet excluded by the experiments within current experimental errors, they are not favoured. Noting the
fact that the very small Higgs-DM coupling leads to a scattering
cross-section that will not be seen at the DM direct detection
experiments, it will be the LHC
that will provide us with the strongest exclusion limits for this 
scenario.

\subsection{{Strong 1st order phase transition, CP violation and bariogenesis}}

A significant difference between the cIDMS and the IDM is a  mixing between the SM-like doublet and a singlet, resulting in 
three Higgs states $h_1$, $h_2$ and $h_3$. Note, {however}, that all three Higgs particles couple to vector boson pairs (e.g. $h_i ZZ$, $h_iW^+W^-$) in a similar manner, with {a} reduced strenght {compared to} the SM- Higgs particles. 
 LHC constraints {for the 125 \GeV~Higgs boson coupling to gauge bosons  being close to the SM value} {(e.g. for ZZ the 1-$\sigma$ interval is  0.92-1.00)
\cite{Khachatryan:2016vau})}, {if applied {to the} $h_1ZZ$ {within the cIDMS}}, {leads to small} corresponding couplings of $h_2$ and $h_3$, as they fulfill corresponding sum rules {\cite{Gunion:1990kf}}.
 
In principle the vacuum state with {a} non-zero phase for a singlet {can induce CP violation}.  
CP violating vacuum and possible  CP-violation in the scalar sector are due to  the {singlet} cubic terms  in the potential, {which are} proportional to parameters $\kappa_2$ or $\kappa_3$, as shown in \cite{Darvishi:2016gvm}. 
These terms are also crucial to obtain strong enough first order phase transition already in the high temperature approximation.  Note, that for IDM these conditions were obtained using the one-loop effective potential approach \cite{Hambye:2007vf,Chowdhury:2011ga,Gil:2012ya}. {It suggests that the cIDMS can lead to {an} even stronger first order phase transition, {as it contains} two sources {for it}. }

{For parameter {regions} {providing {a} sufficiently strong} first-order phase transition}, the next step is to check {whether the} correct baryon asymmetry in the Universe can be generated {in the model}. It is known that a Standard Model with a complex singlet can provide   a strong enough first-order phase transition \cite{Darvishi:2016tni}  and  enough CP violation if an iso-doublet vector quark is added to the model,
since {in this case} the Yukawa Lagrangian acquires additional terms, which depend on the CP-violating phases. More studies are needed to establish whenever it is also possible to have a stable DM particle at the same time, in agreement with all experimental constraints. If it is indeed the case, the cIDMS would be a model that could address these important problems, while being testable at the LHC.

\section{Conclusions and Outlook}
{In this note, several extensions of the SM scalar sector were discussed. These extensions are based on the Inert Doublet Model, a two-Higgs doublet model with a discrete $Z_2$ symmetry rendering a dark matter candidate. First,}
we  presented an update of previous work \cite{Ilnicka:2015jba}, where we thoroughly investigated the parameter space of the Inert Doublet Model. Here, we include more recent constraints on signal strength for 125 \GeV\, Higgs from a combined ATLAS and CMS analysis \cite{Khachatryan:2016vau}, as well as updates on direct detection for dark matter candidates from the LUX experiment \cite{Akerib:2016vxi}. We show how these modify our previous work, and lead to relatively large restrictions in the models parameter space, partially decreasing the available production cross sections by at least one order of magnitude. 

We also discuss an extension of the IDM, {denoted} the cIDMS, where a complex
scalar field with a complex VEV is added to the model. It shares a lot of
features of the IDM, but the presence of additional Higgs particles can
change DM annihilation scenarios, opening new parameter regions. By
introducing an additional source of CP violation we make a step towards
explaining the origin of baryon asymmetry of the Universe -- a feature
that is missing from the IDM.


\begin{thebibliography}{10}

\bibitem{moriond}
Recontres de Moriond 2017.

\bibitem{Deshpande:1977rw}
Nilendra~G. Deshpande and Ernest Ma.
\newblock {Pattern of Symmetry Breaking with Two Higgs Doublets}.
\newblock {\em Phys.Rev.}, D18:2574, 1978.

\bibitem{Cao:2007rm}
Qing-Hong Cao, Ernest Ma, and G.~Rajasekaran.
\newblock {Observing the Dark Scalar Doublet and its Impact on the
  Standard-Model Higgs Boson at Colliders}.
\newblock {\em Phys.Rev.}, D76:095011, 2007, 0708.2939.

\bibitem{Barbieri:2006dq}
Riccardo Barbieri, Lawrence~J. Hall, and Vyacheslav~S. Rychkov.
\newblock {Improved naturalness with a heavy Higgs: An Alternative road to LHC
  physics}.
\newblock {\em Phys.Rev.}, D74:015007, 2006, hep-ph/0603188.

\bibitem{Dolle:2009ft}
Ethan Dolle, Xinyu Miao, Shufang Su, and Brooks Thomas.
\newblock {Dilepton Signals in the Inert Doublet Model}.
\newblock {\em Phys.Rev.}, D81:035003, 2010, 0909.3094.

\bibitem{Gustafsson:2012aj}
Michael Gustafsson, Sara Rydbeck, Laura Lopez-Honorez, and Erik Lundstrom.
\newblock {Status of the Inert Doublet Model and the Role of multileptons at
  the LHC}.
\newblock {\em Phys.Rev.}, D86:075019, 2012, 1206.6316.

\bibitem{Blinov:2015qva}
Nikita Blinov, Jonathan Kozaczuk, David~E. Morrissey, and Alejandro de~la
  Puente.
\newblock {Compressing the Inert Doublet Model}.
\newblock {\em Phys. Rev.}, D93(3):035020, 2016, 1510.08069.

\bibitem{Ferreira:2015pfi}
P.~M. Ferreira and Bogumila Swiezewska.
\newblock {One-loop contributions to neutral minima in the inert doublet
  model}.
\newblock {\em JHEP}, 04:099, 2016, 1511.02879.

\bibitem{Diaz:2015pyv}
Marco~Aurelio Diaz, Benjamin Koch, and Sebastian Urrutia-Quiroga.
\newblock {Constraints to Dark Matter from Inert Higgs Doublet Model}.
\newblock {\em Adv. High Energy Phys.}, 2016:8278375, 2016, 1511.04429.

\bibitem{Hashemi:2015swh}
Majid Hashemi, Maria Krawczyk, Saereh Najjari, and Aleksander~F. Żarnecki.
\newblock {Production of Inert Scalars at the high energy $e^+ e^-$ colliders}.
\newblock 2015, 1512.01175.
\newblock [JHEP02,187(2016)].

\bibitem{Krawczyk:2015xhl}
Maria Krawczyk, Neda Darvishi, and Dorota Sokolowska.
\newblock {The Inert Doublet Model and its extensions}.
\newblock {\em Acta Phys. Polon.}, B47:183, 2016, 1512.06437.

\bibitem{Poulose:2016lvz}
P.~Poulose, Shibananda Sahoo, and K.~Sridhar.
\newblock {Exploring the Inert Doublet Model through the dijet plus missing
  transverse energy channel at the LHC}.
\newblock {\em Phys. Lett.}, B765:300--306, 2017, 1604.03045.

\bibitem{Kanemura:2016sos}
Shinya Kanemura, Mariko Kikuchi, and Kodai Sakurai.
\newblock {Testing the dark matter scenario in the inert doublet model by
  future precision measurements of the Higgs boson couplings}.
\newblock {\em Phys. Rev.}, D94(11):115011, 2016, 1605.08520.

\bibitem{Datta:2016nfz}
Amitava Datta, Nabanita Ganguly, Najimuddin Khan, and Subhendu Rakshit.
\newblock {Exploring collider signatures of the inert Higgs doublet model}.
\newblock {\em Phys. Rev.}, D95(1):015017, 2017, 1610.00648.

\bibitem{deFlorian:2016spz}
D.~de~Florian et~al.
\newblock {Handbook of LHC Higgs Cross Sections: 4. Deciphering the Nature of
  the Higgs Sector}.
\newblock 2016, 1610.07922.

\bibitem{Hashemi:2016wup}
Majid Hashemi and Saereh Najjari.
\newblock {Observability of Inert Scalars at the LHC}.
\newblock 2016, 1611.07827.

\bibitem{Belyaev:2016lok}
Alexander Belyaev, Giacomo Cacciapaglia, Igor~P. Ivanov, Felipe Rojas, and Marc
  Thomas.
\newblock {Anatomy of the Inert Two Higgs Doublet Model in the light of the LHC
  and non-LHC Dark Matter Searches}.
\newblock 2016, 1612.00511.

\bibitem{Ginzburg:2010wa}
I.F. Ginzburg, K.A. Kanishev, M.~Krawczyk, and D.~Sokolowska.
\newblock {Evolution of Universe to the present inert phase}.
\newblock {\em Phys.Rev.}, D82:123533, 2010, 1009.4593.

\bibitem{Hambye:2007vf}
Thomas Hambye and Michel~H.G. Tytgat.
\newblock {Electroweak symmetry breaking induced by dark matter}.
\newblock {\em Phys.Lett.}, B659:651--655, 2008, 0707.0633.

\bibitem{Chowdhury:2011ga}
Talal~Ahmed Chowdhury, Miha Nemevsek, Goran Senjanovic, and Yue Zhang.
\newblock {Dark Matter as the Trigger of Strong Electroweak Phase Transition}.
\newblock {\em JCAP}, 1202:029, 2012, 1110.5334.

\bibitem{Gil:2012ya}
Grzegorz Gil, Piotr Chankowski, and Maria Krawczyk.
\newblock {Inert Dark Matter and Strong Electroweak Phase Transition}.
\newblock {\em Phys.Lett.}, B717:396--402, 2012, 1207.0084.

\bibitem{Goudelis:2013uca}
A.~Goudelis, B.~Herrmann, and O.~Stal.
\newblock {Dark matter in the Inert Doublet Model after the discovery of a
  Higgs-like boson at the LHC}.
\newblock {\em JHEP}, 1309:106, 2013, 1303.3010.

\bibitem{Swiezewska:2015paa}
Bogumila Swiezewska.
\newblock {Inert scalars and vacuum metastability around the electroweak
  scale}.
\newblock {\em JHEP}, 07:118, 2015, 1503.07078.

\bibitem{Khan:2015ipa}
Najimuddin Khan and Subhendu Rakshit.
\newblock {Constraints on inert dark matter from the metastability of the
  electroweak vacuum}.
\newblock {\em Phys. Rev.}, D92:055006, 2015, 1503.03085.

\bibitem{LopezHonorez:2006gr}
Laura Lopez~Honorez, Emmanuel Nezri, Josep~F. Oliver, and Michel~H.G. Tytgat.
\newblock {The Inert Doublet Model: An Archetype for Dark Matter}.
\newblock {\em JCAP}, 0702:028, 2007, hep-ph/0612275.

\bibitem{Ilnicka:2015jba}
Agnieszka Ilnicka, Maria Krawczyk, and Tania Robens.
\newblock {Inert Doublet Model in light of LHC Run I and astrophysical data}.
\newblock {\em Phys. Rev.}, D93(5):055026, 2016, 1508.01671.

\bibitem{Ilnicka:2015sra}
Agnieszka Ilnicka, Maria Krawczyk, and Tania Robens.
\newblock {Constraining the Inert Doublet Model}.
\newblock In {\em {2nd Toyama International Workshop on Higgs as a Probe of New
  Physics (HPNP2015) Toyama, Japan, February 11-15, 2015}}, 2015, 1505.04734.

\bibitem{Ilnicka:2015ova}
Agnieszka~Janina Ilnicka, Maria Krawczyk, and Tania Robens.
\newblock {Inert Doublet Model in the light of LHC and astrophysical data}.
\newblock {\em PoS}, EPS-HEP2015:143, 2015, 1510.04159.

\bibitem{Bonilla:2014xba}
Cesar Bonilla, Dorota Sokolowska, Neda Darvishi, J.~Lorenzo Diaz-Cruz, and
  Maria Krawczyk.
\newblock {IDMS: Inert Dark Matter Model with a complex singlet}.
\newblock {\em J. Phys.}, G43(6):065001, 2016, 1412.8730.

\bibitem{Chuzhoy:2008zy}
Leonid Chuzhoy and Edward~W. Kolb.
\newblock {Reopening the window on charged dark matter}.
\newblock {\em JCAP}, 0907:014, 2009, 0809.0436.

\bibitem{Khachatryan:2016ctc}
Vardan Khachatryan et~al.
\newblock {Search for Higgs boson off-shell production in proton-proton
  collisions at 7 and 8 TeV and derivation of constraints on its total decay
  width}.
\newblock {\em JHEP}, 09:051, 2016, 1605.02329.

\bibitem{Khachatryan:2016vau}
Georges Aad et~al.
\newblock {Measurements of the Higgs boson production and decay rates and
  constraints on its couplings from a combined ATLAS and CMS analysis of the
  LHC pp collision data at $ \sqrt{s}=7 $ and 8 TeV}.
\newblock {\em JHEP}, 08:045, 2016, 1606.02266.

\bibitem{Akerib:2016vxi}
D.~S. Akerib et~al.
\newblock {Results from a search for dark matter in the complete LUX exposure}.
\newblock {\em Phys. Rev. Lett.}, 118(2):021303, 2017, 1608.07648.

\bibitem{Bechtle:2008jh}
Philip Bechtle, Oliver Brein, Sven Heinemeyer, Georg Weiglein, and Karina~E.
  Williams.
\newblock {HiggsBounds: Confronting Arbitrary Higgs Sectors with Exclusion
  Bounds from LEP and the Tevatron}.
\newblock {\em Comput.Phys.Commun.}, 181:138--167, 2010, 0811.4169.

\bibitem{Bechtle:2011sb}
Philip Bechtle, Oliver Brein, Sven Heinemeyer, Georg Weiglein, and Karina~E.
  Williams.
\newblock {HiggsBounds 2.0.0: Confronting Neutral and Charged Higgs Sector
  Predictions with Exclusion Bounds from LEP and the Tevatron}.
\newblock {\em Comput.Phys.Commun.}, 182:2605--2631, 2011, 1102.1898.

\bibitem{Bechtle:2013wla}
Philip Bechtle, Oliver Brein, Sven Heinemeyer, Oscar Stal, Tim Stefaniak,
  et~al.
\newblock {$\mathsf{HiggsBounds}-4$: Improved Tests of Extended Higgs Sectors
  against Exclusion Bounds from LEP, the Tevatron and the LHC}.
\newblock {\em Eur.Phys.J.}, C74(3):2693, 2014, 1311.0055.

\bibitem{Bechtle:2013xfa}
Philip Bechtle, Sven Heinemeyer, Oscar Stal, Tim Stefaniak, and Georg Weiglein.
\newblock {$HiggsSignals$: Confronting arbitrary Higgs sectors with
  measurements at the Tevatron and the LHC}.
\newblock {\em Eur.Phys.J.}, C74(2):2711, 2014, 1305.1933.

\bibitem{Akerib:2013tjd}
D.S. Akerib et~al.
\newblock {First results from the LUX dark matter experiment at the Sanford
  Underground Research Facility}.
\newblock {\em Phys.Rev.Lett.}, 112(9):091303, 2014, 1310.8214.

\bibitem{Planck:2015xua}
P.A.R. Ade et~al.
\newblock {Planck 2015 results. XIII. Cosmological parameters}.
\newblock 2015, 1502.01589.

\bibitem{Krawczyk:2015vka}
Maria Krawczyk, Malgorzata Matej, Dorota Sokolowska, and Bogumila Swiezewska.
\newblock {The Universe in the Light of LHC}.
\newblock {\em Acta Phys. Polon.}, B46(1):169--179, 2015, 1501.04529.

\bibitem{Aprile:2012zx}
Elena Aprile.
\newblock {The XENON1T Dark Matter Search Experiment}.
\newblock {\em Springer Proc. Phys.}, 148:93--96, 2013, 1206.6288.

\bibitem{Gunion:1990kf}
J.~F. Gunion, H.~E. Haber, and J.~Wudka.
\newblock {Sum rules for Higgs bosons}.
\newblock {\em Phys. Rev.}, D43:904--912, 1991.

\bibitem{Darvishi:2016gvm}
Neda Darvishi and Maria Krawczyk.
\newblock {CP violation in the Standard Model with a complex singlet}.
\newblock 2016, 1603.00598.

\bibitem{Darvishi:2016tni}
Neda Darvishi.
\newblock {Baryogenesis of the Universe in cSMCS Model plus Iso-Doublet Vector
  Quark}.
\newblock {\em JHEP}, 11:065, 2016, 1608.02820.

\end{thebibliography}

\end{document}